# A numerical model for the study of photoacoustic imaging of brain tumours

**Kamyar Firouzi[1] and Nader Saffari[2]**

[1]Department of Mechanical Engineering, Stanford University, Stanford, CA
[2]Department of Mechanical Engineering, University College London, UK

Email: kfirouzi@stanford.edu

**Abstract.** Photoacoustic imaging has shown great promise for medical imaging, where optical energy absorption by blood haemoglobin is used as the contrast mechanism. A numerical method was developed for the *in-silico* assessment of the photoacoustic image reconstruction of the brain. Image segmentation techniques were used to prepare a digital phantom from MR images. Light transport through brain tissue was modelled using a Finite Element approach. The resulting acoustic pressure was then estimated by pulsed photoacoustics considerations. The forward acoustic wave propagation was modelled by the linearized coupled first order wave equations and solved by an acoustic *k*-space method. Since skull bone is an elastic solid and strongly attenuates ultrasound (due to both scattering and absorption), a *k*-space method was developed for elastic media. To model scattering effects, a new approach was applied based on propagation in random media. In addition, absorption effects were incorporated using a power law. Finally, the acoustic pressure was reconstructed using the *k*-space time reversal technique. The simulations were ran in 3D to produce the photoacoustic tomogram of a brain tumour. The results demonstrate the convergence of the models, and their suitability for investigating the photoacoustic imaging process.

## 1. Introduction

One of the new techniques for biomedical imaging is a modality called photoacoustic (PA) imaging, which combines high optical contrast and high ultrasound resolution. PA can overcome the disadvantage of optical imaging in spatial resolution and the disadvantage of ultrasound imaging in contrast and speckle artefact. It is based on the photoacoustic effect, a phenomenon in which the absorbed energy from light is transformed into the kinetic energy of the sample by energy exchange processes, leading to local heating and thus a pressure wave. In other words, the PA effect is the generation of acoustic waves by the absorption of electromagnetic energy such as optical or radio-frequency waves. Often, when visible light or near infrared (NIR) wavelengths are applied, it is referred as the *photoacoustic* effect, and when radio-frequency waves are used, it is called the *thermoacoustic* effect.

In biomedical photoacoustic imaging, pulsed laser light is used at wavelengths which are predominantly absorbed by haemoglobin in blood. This causes a small local rise in temperature, restricted dilatation of the red blood cells and subsequent elevation of the local pressure. Consequently, ultrasonic waves are generated, which, when measured in a sufficient number of locations at the tissue surface, can be used to reconstruct the position of blood vessels (Siphanto *et al* 2005). The optical absorption of whole blood is much stronger than that of other tissues. Therefore, blood generates strong photoacoustic signals and manifests high image contrast, causing the vasculature in organs to stand out prominently in photoacoustic images (Ku *et al* 2005).

The application of photoacoustics for detection of tumours has also been of recent interest. It has shown great promise for visualising different anatomical organs and especially the vasculature in human body. It has also been used for imaging certain types of tumours, such as that of the breast (Pramanik *et al* 2008, Jose *et al* 2009). However, it has not as yet been applied for imaging tumours of brain. although there have been some limited investigations on animal models such as the rat (Siphanto *et al* 2005, Ku *et al* 2005, Li *et al* 2008) and ape (Yang and Wang 2006, 2008). These results point to the success of the method in animal models but, in the case of human brain tumours, its feasibility has remained largely unproven.

The aim of this study was to evaluate the feasibility of using photoacoustic imaging for the detection of brain tumours. One of the drivers behind this work has been the need to find an alternative to MR (Magnetic Resonance) imaging. In MR imaging of brain tumours, Gadolinium is generally used as a contrast agent. There is growing evidence however of the toxicity of Gadolinium particularly in the case of some groups of susceptible patients with poor renal function (Perazella 2008).

The hypothesis behind the work in this paper is that brain tumours are either hyper-vascular or have a hypo-vascular necrotic core. Since light absorption by blood haemoglobin is the main contrast mechanism, we will have a relative change in image intensity at the location of the tumour and hence the tumour will present itself with a reasonable contrast against the surrounding healthy tissue.

To conduct a detailed study of the efficacy of photoacoustic imaging using either human or animal subjects *in-vivo* is not a practical proposition. *Ex-vivo* animal models do not provide a good representation of human brain tumours, because of the differences in the structure of the skull bone. On the other hand, fabrication of detailed, vascular phantoms is difficult and expensive. Therefore, the emphasis of the work has been on developing a new method for a detailed *in-silico* (simulation-based) assessment of the above hypothesis by simulating the entire photoacoustic image reconstruction process of the brain. This requires the development of appropriate techniques for modelling of the optical and acoustic fields in soft tissues (i.e. the brain tissues). Whereas the brain itself can be considered to be a fluid medium and does not support shear stresses, the skull bone is an elastic solid and as such is an effective shear wave propagator and strongly attenuates any ultrasound (due to both scattering and absorption). This necessitates the development of a new modelling method which can correctly account for shear stresses and accurately estimate the attenuation effects. Moreover, a suitable reconstruction algorithm is of interest for the completion of the imaging process.

## 2. Methodology

### 2.1. Preparation of the in-silico phantom

*2.1.1. Requirements.* For the simulation of both optical and acoustic fields, it is essential to accurately define the photoacoustic sources (i.e. within the vasculature) and the photoacoustic medium (i.e. brain tissue) characteristics. The Brain is a highly heterogonous and thus a highly scattering medium for both light diffusion and ultrasound propagation; therefore, it cannot be considered as a uniform medium. Besides, as stated earlier, at the typical IR wavelengths used in photoacoustic imaging, the main absorber of light is blood haemoglobin and thus a knowledge of the vascular tree is also essential.

The *in-silico* phantoms used in this study, which provided the precise anatomical detail of the brain were in the form of Gadolinium-enhanced, T1, T2 weighted MR images. For vasculature segmentation, as blood vessels are not diagnosable in T1 and T2 weighted MR, magnetic resonance angiography (MRA) images were used. In angiograms, blood appears in the brightest colours. Precise definition of brain and tumour structure and morphology is feasible by applying suitable image processing techniques, including segmentation and registration methods. Therefore, we segmented the brain images prior to simulating the photoacoustic field. The different segmented tissue types are skull, brain white matter, gray matter, CSF (cerebrospinal fluid), coupling medium (to be considered as water), edema, tumour, and the vasculature.

*2.1.2. Brain and tumour tissues segmentation.* Segmentation of the brain images (excluding the vasculature) was implemented using an EM (Expectation and Maximization) algorithm. For this purpose, 3DSlicer (Gering *et al* 1999, Pieper *et al* 2004, 2006) software package, which has suitable functionality for EM segmentation, was used. EM is almost an automatic segmentation method; however, some manual adjustments are necessary to achieve a good level of segmentation, as will be explained later. EM is a statistical ATLAS-based method. An ATLAS for this method of segmentation would be in the form of a probability image (i.e. a so called ground truth or spatial prior), which is tissue-specific, and the voxel intensities demonstrate the likelihood of finding a specific tissue in that particular voxel.

This method is suitable for segmenting tissues of a healthy brain; however, there is a challenge in segmentation of a cancerous brain using this approach. Automated tumour segmentation is difficult because one cannot always create a prior model of expected size, shape, location, or image intensity. When a healthy brain ATLAS is used as a spatial prior for tissue classification, all brain tissue must be classified as one of the available tissue types, i.e. gray matter, white matter, and CSF. Tumours consist of two tissue types (tumour and edema), not present in the ATLAS of normal patients. Moreover, it may additionally change the image intensities of normal anatomical structures via infiltration or edema (Prastawa *et al* 2009). This implies the amount and the regional extent of edema that accompanies a tumour could be variable. Thus, a normal brain ATLAS cannot be used when applying the EM method.

To overcome this problem, we used the method developed by Moon *et al* (2002) and Prastawa *et al* (2003, 2005, 2009), which is an extended EM algorithm to be applied for the case of cancerous brain segmentation. This method provides appropriate synthetic probabilistic images of a cancerous brain from a healthy brain ground truth. The process is started by manually defining a seed point; probabilistic images are then produced with respect to the mass effect and infiltration of the predicted brain tumour. Also, two new pathological probabilities (tumour and edema) are added to the previous probabilities (gray matter, white matter, and CSF). In the following, figure 1 shows the original MR and figure 2 illustrates the segmented image, where each tissue class of a cancerous brain is labelled with a specific colour to be easily distinguished from its surrounding tissues.

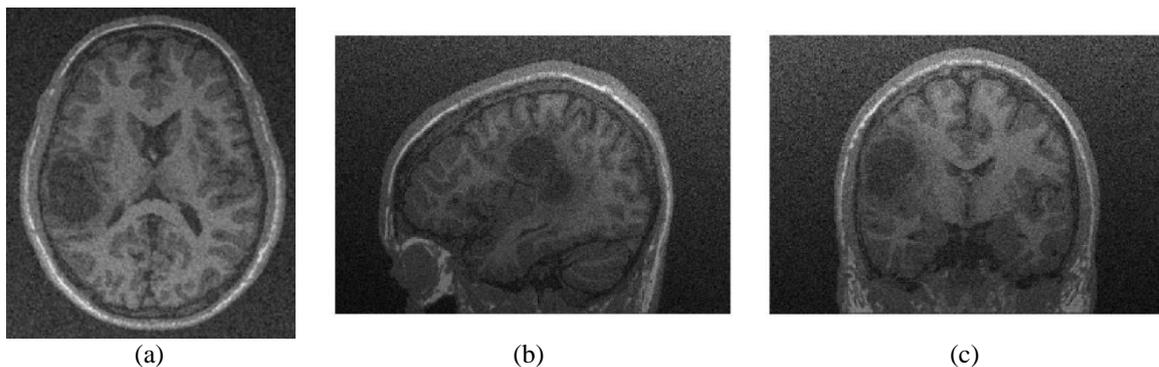

(a)　　　　　　　　　　　　(b)　　　　　　　　　　　　(c)
**Figure 1.** Original T1 weighted MR of a cancerous brain obtained from (http://www.ucnia.org/softwaredata/5-tumordata/10-simtumordb.html). (a) Axial. (b) Sagittal. (c) Coronal.

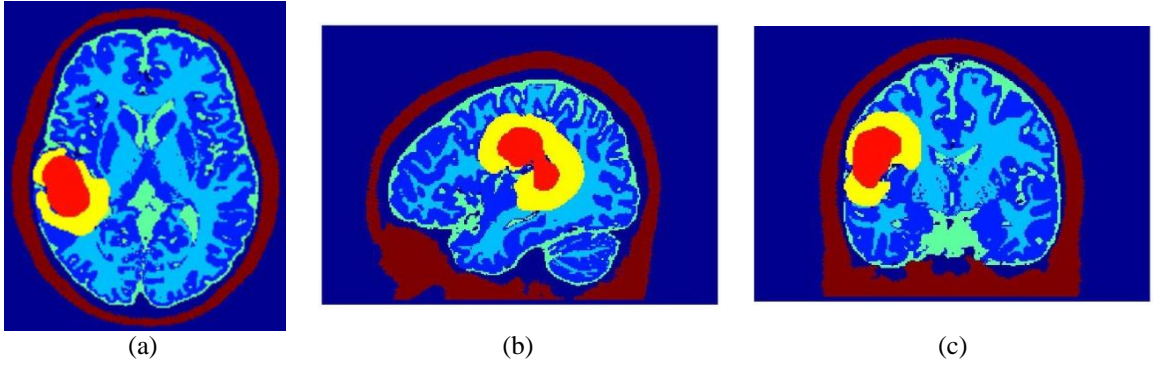
**Figure 2.** Segmented brain and tumour tissues. (a) Axial. (b) Sagittal. (c) Coronal.

The process of the tissue segmentation of the cancerous brain using 3D Slicer is shown in figure 3.

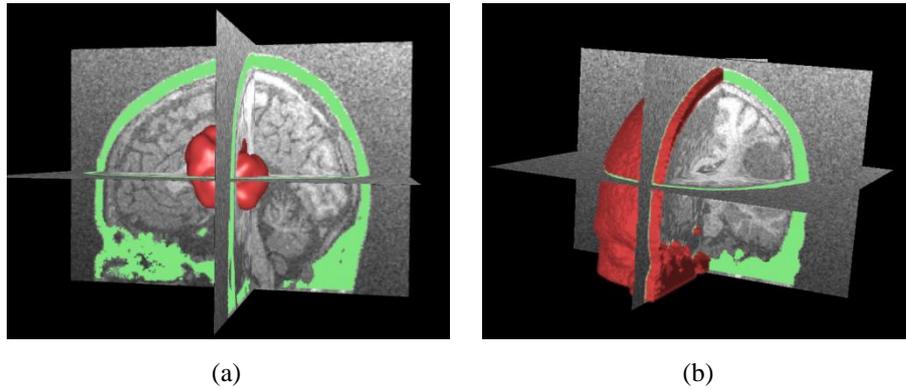
**Figure 3.** Process of segmentation of the skull and the tumour. (a) Tumour segmentation. (b) Skull segmentation.

*2.1.3. Brain Vasculature segmentation.* The brain vasculature were segmented using a seed-point-based level-set algorithm (Antiga *et al* 2008). In this work, there is a need for such segmentation algorithm since the cerebral blood volume (CBV) is the main optical absorber and and therefore contains the acoustic sources of the *in-silico* model. Here, the Vascular Modelling Toolkit (VMTK) (Antiga *et al* 2008, Hähn 2009), which provides a collection of libraries for image based modelling of blood vessels, incorporated within 3DSlicer, was applied to segment the brain and tumour vasculature. The segmentation was implemented in three stages; first, the blood contrast in MRA was enhanced using a filtering algorithm, then the level-set segmentation was applied, and finally the vessels centrelines were extracted and vessel trees were formed. The results are presented in figure 4.

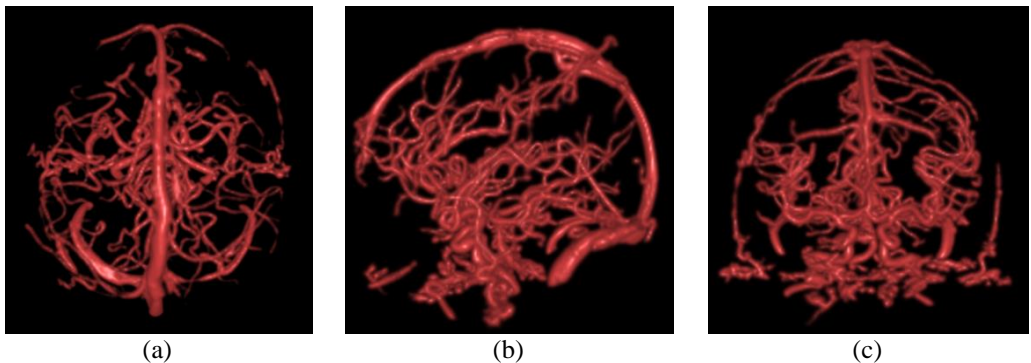
**Figure 4.** Segmented vasculature (slice views). (a) Axial. (b) Sagittal. (c) Coronal.

*2.1.4. Registration.* The heterogeneity in tissues optical properties affects the diffused and absorbed light in tissues (i.e. mainly scattering effects).Therefore in modelling light transport through tissue, it is essential to use a mask file containing both the segmented brain tissues and the segmented vasculature. So, the results of the previous sections were registered and mapped onto each other using affine registration to provide an assembled image of both the segmented vasculature and tissues. However, for simulating the acoustic field, only the segmented brain and tumour tissues affect wave propagation and the vasculature have a negligible effect. Hence, for the purpose of defining the heterogeneity in the acoustic field, we only employed the results of section 2.1.2.

*2.2. Numerical modelling of the photoacoustic field*

*2.2.1. Light transport through tissue.* To derive the first half of the forward photoacoustic problem, a model for photon transport in tissue is required. There are two basic approaches (Schweiger *et al* 1993b), an essentially discrete model of individual photon interactions, such as a Monte Carlo method (van der Zee 1992), or a continuous model based on a differential equation approximation (Schweiger 1994), such as the diffusion equation. Both of these approaches are derived from an integro-differential equation description. This is written as (Schweiger *et al* 1992, 1993b),

$$\left\{\hat{s}\cdot\nabla+\mu_a+\mu_s+\frac{1}{v}\frac{\partial}{\partial t}\right\}\phi(r,\hat{s}',t)=\frac{1}{v}q(\hat{s},\hat{r},t)+\mu_s\int f(\hat{s}',\hat{s})\phi(r,\hat{s}',t)\,d^2\hat{s} \qquad (1)$$

where $\mu_a$, $\mu_s$ are spatially varying absorption and scattering coefficients, respectively, $\hat{r},\hat{s},\hat{s}'$ are direction vectors, $v$ is the speed of light in the medium, $\phi(r,\hat{s}',t)$ is the photon density in direction $\hat{s}'$, $q$ is the source term, and $f(\hat{s}',\hat{s})$ is the probability of scattering from direction $\hat{s}'$, into direction $\hat{s}$. Photon diffusion approximation is a second order partial differential equation describing the time behaviour of photon fluence rate distribution in a low-absorption high-scattering medium (Schweiger *et al* 1993b). Its main difference with diffusion equation in physics is that the photon diffusion equation has an absorption term in it. Its mathematical form is as follows.

$$\left\{\nabla\cdot\kappa\nabla-\mu_a c-\frac{\partial}{\partial t}\right\}\Phi(r,t)=-q_0(r,t) \qquad (2)$$

where $\Phi$ is photon fluence rate and $\kappa(r)$ is the diffusion coefficient given as,

$$\kappa=\frac{v}{3(\mu_a+(1-g)\mu_s)} \qquad (3)$$

Here, $g$ is the average cosine of the scattering angle distribution, $q_0$ is the source term, and

$$\Phi(r,t)=\int\phi(r,\hat{s}',t)\,d^2\hat{s} \qquad (4)$$

The term $(1-g)\mu_s$ is referred as the reduced scattering coefficient ($\mu_s'$). Experimental and theoretical work has demonstrated the validity of equation (2) under conditions in which $\mu_a\ll\mu_s$, which is the case for the near infrared (NIR) transillumination of tissue, where typical parameter ranges are $0.01<\mu_a<0.1\,mm^{-1}$ and $1.0<\mu_s'<10.0\,mm^{-1}$. For certain applications such as transilluminating very thin tissue layers and time-gated techniques, it may not be scattering dominated and therefore cannot correctly be described by the diffusion approximation (Arridge *et al* 1992, Schweiger *et al* 1993a, b).

*2.3.2. Acoustic wave modelling.* If a region of a fluid is heated through the absorption of a laser pulse, a sound wave is generated. In a stationary fluid with isotropic acoustic properties, under conditions whereby the sound generation mechanism is thermoelastic and the effects of viscosity and thermal conductivity could be ignored, the wave equation for the acoustic pressure is (Cox and Beard 2005, Cox *et al* 2005),

$$\left\{\nabla^2-\frac{1}{c^2}\frac{\partial}{\partial t^2}\right\}p=-\frac{\beta}{C_p}\frac{\partial\mathcal{H}}{\partial t} \qquad (5)$$

where $c$ is the sound speed, $\beta$ is the volume thermal expansivity, $C_p$ is the constant pressure specific heat capacity, $p$ is the acoustic pressure and $\mathcal{H}$ is the heat energy per unit volume and per unit time deposited in the fluid. $p$ and $\mathcal{H}$ will depend, in general, on the position $\mathrm{r} = (x, y, z)$ and time $t$.

*2.2.3. Photoacoustic interaction in tissue.* In biomedical photoacoustic imaging, the spatially-varying pressure increases following the absorption of a laser pulse. This is due to the photoacoustic interaction inside the tissues, in which the absorbed light energy causes a local increase in temperature. Tissues will thus expand and contract rapidly. Consequently, acoustic pressure is generated. From a thermodynamic point of view, the whole process could be formulated using the thermodynamic energy relation and the equation of state as below (Cox and Beard 2005),

$$\rho C_p \frac{\partial \tau}{\partial t} = \mathcal{H} \tag{6}$$

$$p = \frac{1}{\kappa_T}\left(\frac{\delta}{\rho} + \beta\tau\right) \tag{7}$$

where, $\kappa_T$ is the isothermal compressibility, $\rho$ is density, and $\tau$ is change in temperature. In pulsed biomedical photoacoustics, the light pulse is so short that the heating of the absorber occurs instantaneously without expansion (i.e. adiabatic heating), resulting in a pressure rise. When the laser pulse is effectively instantaneous the heating function can be modelled as (Cox and Beard 2005),

$$\mathcal{H}(\mathrm{r}, t) = H(\mathrm{r})\delta_D(t) \tag{8}$$

where $H(\mathrm{r})$ is the heat deposited in the fluid per unit volume (also called the absorbed energy map), and $\delta_D(t)$ is the Dirac delta. This condition is referred as stress confinement and is when the pulse is much shorter than the time it would take for sound to travel across the heated region (Paltauf and Dyer 2003); then, it may be considered as instantaneous and $p_0(\mathrm{r})$ is the initial pressure distribution at $t = 0$. This condition requires that the laser pulse duration $t_p$ (Cox *et al* 2004),

$$t_p \ll \frac{1}{\mu_a c} \tag{9}$$

which implies the optical energy will be absorbed before the fluid density has time to change. Here $\mu_a$ is the optical absorption coefficient of the medium. $1/\mu_a$, which is the optical penetration depth, is referred to as the characteristic length of the heated region. When this condition is met, the spatial part of the heating function can be written as (Cox *et al* 2005)

$$H(\mathrm{r}) = \mu_a(\mathrm{r})\Phi(\mathrm{r}, \mu_a) \tag{10}$$

The fluence $\Phi$ will in general depend on the absorption coefficient distribution $\mu_a(\mathrm{r})$ and the scattering coefficient distribution $\mu_s(\mathrm{r})$, and hence, the absorbed energy $H$ is nonlinearly related to $\mu_a$. In this case, the acoustic pressure immediately following the pulse, or initial pressure distribution, $p_0(r)$, is proportional to the absorbed energy map.

$$p_0(\mathrm{r}) = \left(\frac{\beta c^2}{C_p}\right)H(\mathrm{r}) = \Gamma H(\mathrm{r}) = \Gamma\mu_a(\mathrm{r})\Phi(\mathrm{r}, \mu_a) \tag{11}$$

$\Gamma$ is called the Grüneisen coefficient, a dimensionless constant that represents the efficiency of the conversion of heat to pressure. This initial pressure distribution then propagates away as acoustic waves according to equation (5). This means that the problem can be recast as an initial value problem, i.e. with no explicit acoustic sources but an initial value for the acoustic pressure. In other words the photoacoustic wave equation could be rewritten as (Cox *et al* 2007a),

$$\left\{\nabla^2 - \frac{1}{c^2}\frac{\partial}{\partial t^2}\right\}p = 0, \quad p|_{t=0} = p_0(\mathrm{r}), \quad \left.\frac{\partial p}{\partial t}\right|_{t=0} = 0 \tag{12}$$

When modelling linear sound propagation in soft biological tissue, it is usually assumed that the propagation medium is isotropic, that there is no net flow, and that shear waves can be neglected. Under these conditions the acoustic pressure, $p$, acoustic particle velocity, $u$, and acoustic density, $\delta$, are related by three first order equations corresponding to momentum conservation, mass conservation, and an equation of state (Cox *et al* 2007b).

$$\frac{\partial u}{\partial t} = -\frac{1}{\rho}\nabla p, \quad \frac{\partial \delta}{\partial t} = -\rho \nabla \cdot u, \quad p = c^2 \delta \qquad (13)$$

where the sound speed $c$ and ambient density $\rho$ can vary with position, and the pressure, $p$, will depend, in general, on both position and time.

*2.2.4. Numerical implementation.* For the simulation of photon transport in tissue, equation (2) was numerically solved using the finite element (FE) method developed by Schweiger *et al* (1992, 1993a, b) and Schweiger (1994). This model calculates the integrated intensity and the mean time of flight from the boundary flux for given tissue parameters and source distribution. The advantage of this numerical approach is its flexibility. It can be applied to complex geometries and inhomogeneous parameter distributions. For implementation, we used an in-house MATLAB toolbox called TOAST (Schweiger and Arridge *2008*), which uses the FE approach for solving equation (2) for determining the light transport.

In order to model the acoustic field, a pseudo-spectral and *k*-space approach was used (Tabei *et al* 2002). This approach has significant advantages over the more common finite difference (FD) method. Although FD is excellent for many applications, for time domain modelling of broadband or high-frequency waves, it can become cumbersome and slow. This is due to the need to have many grid points per wavelength and small time-steps in order to minimize unwanted numerical dispersion. The pseudo-spectral (PS) method can help to reduce the former, and the *k*-space method can help to overcome the latter (Cox *et al* 2007b). Pseudo-spectral methods in which the spatial fields are calculated via fast Fourier transforms (FFT) have been proven to be more advantageous than FE or FD since they fit a Fourier series to all the data on each line of the mesh; so spatial derivatives are calculated faster and easier, and also only two nodes per wavelength are required in order to describe a wave, rather than >10 in FE and FD methods (Goursolle *et al* 2007, Cox *et al* 2007b). However, they have been extensively coupled with FD schemes for solving the temporal domain of the problem. This may cause instabilities and dispersion and small time steps are required to minimize them (Cox *et al* 2007b).

A *k*-space approach could be considered as a modified pseudo-spectral technique for solving wave equations. *k*-space modifies the standard differencing method for time integration by introducing a periodic function, so that much larger time steps can be chosen without introducing inaccuracy and instability. Therefore, it results in significant computation time and memory savings compared to other numerical methods (Cox *et al* 2007b).

In classical pseudo-spectral methods, where the time derivatives are calculated using a finite difference scheme, the spatial derivatives are calculated in the spatial frequency domain and transformed back to the real domain using inverse fast Fourier transform (IFFT),

$$\partial[\cdot]/\partial x_j = \text{FFT}^{-1}\left\{ik_{x_j}\text{FFT}\{[\cdot]\}\right\} \qquad (14)$$

where, FFT, FFT$^{-1}$ represent fast Fourier and inverse fast Fourier transforms. $k_{x_j}$ is the wave number in the direction of $x_j$. It would be computationally more effective to use an FFT algorithm for the calculation of the Fourier transform, where only $N\log(N)$ operations are needed instead of $N^2$. In the *k*-space scheme, a spatial derivative operator is modified by a temporal propagator as shown by equation (15),

$$\partial[\cdot]/\partial x_j = \text{FFT}^{-1}\left\{ik_{x_j}sinc\left(\frac{c_\infty k \Delta t}{2}\right)\text{FFT}[\cdot]\right\} \qquad (15)$$

where $c_\infty$ is the maximum sound speed of a heterogeneous medium, $k$ is the total wave number, and $\Delta t$ is a small time-interval. Using equation (15), the time derivatives can then be calculated via a finite difference scheme but this time without instability problems. Full details of the mathematical derivation of equation (15) are given by Tabei *et al* (2002). For the purpose of implementation, we used *k*-Wave (an in-house MATLAB toolbox) (Treeby and Cox 2010a), which solves equation (13) using the discussed *k*-space method.

*2.3. Numerical modelling of the skull bone*

*2.3.1. k-space model for the skull bone as an elastic frame.* In the previous section, a suitable method was explained for modelling the forward acoustic field, where a pseudo-spectral and *k*-space method was applied for discretizing the longitudinal wave propagation in an acoustic medium. This is a suitable model for soft tissues such as brain. However skull bone may act as an elastic medium which can support shear waves as well as longitudinal waves. Here, in order to model the shear wave propagation, the skull was considered as an isotropic homogenous elastic medium.

Moreover, bone has a cancellous structure in the middle. This implies high scattering effects for the incident acoustic waves, and accordingly, large attenuation. Thus, the skull cannot be assumed to behave as an acoustic medium. Inclusion of shear wave propagation in ultrasound models has been widely ignored due to either being of insignificant amplitude or being hard to predict (Clement *et al* 2006). However, according to recent studies, the assumption that the transcranial propagation is composed of purely longitudinal modes is only valid for small incident wave angles (Clement *et al* 2004); this assumption rapidly breaks down when incident waves go beyond Snell's critical angle, whereby there would be a mode conversion from an incident longitudinal wave into a shear wave in the bone layers.

Most previous studies involving modelling the skull have focused on the assessing its effects on ultrasound images, where the acoustic beams are often considered normal with respect to skull (Hayner and Hynynen 2001, Clement *et al* 2006). In photoacoustics however, the incident acoustic waves to the skull surface may be in any arbitrary angle. Therefore, it is essential to model shear wave propagation through the skull as well as longitudinal waves.

For this study, a new *k*-space method for elastic isotropic heterogeneous media was developed in our laboratory, which could then be easily coupled with the method of section 2.2.4 (Firouzi *et al* 2004). The governing equations of the problem for isotropic heterogeneous elastic media are in the form of the following second order equation (Aki and Richards 2002),

$$\rho\, \partial^2 u_i/\partial t^2 = \partial \left[(\lambda + 2\mu)\, \partial u_j/\partial x_j\right]/\partial x_i + \partial \left[\mu\, \partial u_i/\partial x_j\right]/\partial x_j + f_i \qquad (16)$$

where $u_i$ is the displacement component and $f_i$ is the body force. Also, $\rho$, $\lambda$ and $\mu$ are the density and Lame's constants, respectively. For our numerical model of elastic waves, we used the equivalent coupled first order stress-velocity formulation. This is given as (Goursolle *et al* 2007),

$$\partial \sigma_{ij}/\partial t = \lambda \delta_{ij}\, \partial v_k/\partial x_k + \mu\left(\partial v_i/\partial x_j + \partial v_j/\partial x_i\right) \qquad (17a)$$

$$\rho\, \partial v_i/\partial t = \partial \sigma_{ij}/\partial x_j + f_i \qquad (17b)$$

where $v_i$ and $\sigma_{ij}$ show the velocity and stress components. Note that the summation convention is used for equations (16) and (17), and $i,j = x, y, z$, where $x, y, z$ are the spatial coordinates. The new *k*-space model for elastic waves states that each field (velocity and stress components) could be considered as the sum of a compressional and a shear part. Spatial derivatives of each are calculated using equations (18a, b), where the *sinc* functions are the temporal modifiers resulting in much better numerical accuracy and stability than classical pseudo-spectral or finite difference models of elastic waves. Similar to the acoustic *k*-space, the time integrations are then implemented using the standard finite difference scheme.

$$\partial[\cdot]^p/\partial x_j = \mathbf{FFT}^{-1}\left\{ik_{x_j}sinc\left(\frac{c_{p_\infty}k\Delta t}{2}\right)\mathbf{FFT}\{[\cdot]^p\}\right\} \quad (18a)$$

$$\partial[\cdot]^s/\partial x_j = \mathbf{FFT}^{-1}\left\{ik_{x_j}sinc\left(\frac{c_{s_\infty}k\Delta t}{2}\right)\mathbf{FFT}\{[\cdot]^s\}\right\} \quad (18b)$$

where $c_{p_\infty}$ and $c_{s_\infty}$, respectively, denotes the maximum compressional and shear speeds in the elastic medium.

*2.3.2. Scattering effects.* Attenuation effects of the skull need to be accounted for in a realistic and reliable assessment of the photoacoustic imaging of the brain. The attenuation of the skull is due to two important mechanisms, scattering and absorption. The former comes from heterogeneities of the skull bone because of its cancellous structure. Skull, in general, could be considered as a porous medium with a mixture of marrow and bone matrix (Strelitzki *et al* 1998, Nicholson *et al* 2000).

A suitable approach for modelling the scattering effects is to consider the skull as a medium with random variables as the ultrasound characteristics. This method is based on the theory of wave propagation in random media and previously was widely used for deriving and estimating the scattering and the attenuation coefficients of bone (Sehgal and Greenleaf 1984, Strelitzki *et al* 1998, Nicholson *et al* 2000, Guo *et al* 2007). The excellent match produced by this method with the experimental results is rather promising (Strelitzki *et al* 1998). One of the best known models in this area is the binary mixture model of the medium. This was originally proposed by (Strelitzki *et al* 1998) for bone structure. In order to incorporate the scattering effect of the skull in our study, a random binary mixture model was utilised. Our method is based on the generation of high frequency heterogeneities in the homogenous background elastic medium. For this purpose, the total elastic field is subdivided into the background and scattered fields. The heterogeneities are then recast as *induced source terms* onto the background homogenous frame. The induced sources stem from the non-uniformity of the medium characteristics. Non-uniformities are therefore generated using random variable techniques, thus introducing heterogeneities into the numerical model.

Bone has been reported to have a Gaussian autocorrelation function and zero cross correlation for the medium properties (Chernov 1960, Chivers 1977, Strelitzki *et al* 1998). These conditions hold when there is the normal probability distribution of random variables over the medium (Hines *et al* 2003), where field variables are considered as elastic medium properties. For implementation, high frequency distributions of parameters were generated using the estimated mean value and variances of each parameter. The mean value was considered as the estimation provided by the binary mixture model and variances are adjusted by trial and error, to fit the fluctuations between the min and max values of the biphasic mixture. The binary mixture model estimates the average and mean fluctuations of each property based on the porosity of the bone. These are written as,

$$\bar{\rho} = \sum \emptyset_i \rho_i = (1-\emptyset)\rho_c + \emptyset \rho_m \quad (19a)$$

$$\langle \rho^2 \rangle = \emptyset(1-\emptyset)\left(1-\emptyset+\emptyset\left(\frac{\rho_b}{\rho_m}\right)^2\right)\left(\frac{\rho_c-\rho_m}{\rho_m}\right)^2 \quad (19b)$$

$$\bar{c}_p^{-1} = \sum \emptyset_i c_{pi}^{-1} = (1-\emptyset)c_{pc}^{-1} + \emptyset c_{pm}^{-1} \quad (19c)$$

$$\langle \bar{c}_p^2 \rangle = \emptyset(1-\emptyset)\left(1-\emptyset+\emptyset\left(\frac{c_{pc}}{c_{pm}}\right)^2\right)\left(\frac{c_{pc}-c_{pm}}{c_{Pm}}\right)^2 \quad (19d)$$

$$\bar{c}_s^{-1} = \sum \emptyset_i c_{si}^{-1} = (1-\emptyset)c_{sc}^{-1} + \emptyset c_{sm}^{-1} \quad (19e)$$

$$\langle \bar{c}_s^2 \rangle = \emptyset(1-\emptyset)\left(1-\emptyset+\emptyset\left(\frac{c_{sc}}{c_{sm}}\right)^2\right)\left(\frac{c_{sc}-c_{sm}}{c_{sm}}\right)^2 \quad (19f)$$

where, $\emptyset$ is the porosity, or volume fraction of marrow in bone cortical matrix, $\bar{\rho}, \bar{c}_p, \bar{c}_s$ are the known average values, and $\langle \rho^2 \rangle, \langle \bar{c}_p^2 \rangle, \langle \bar{c}_s^2 \rangle$ are the required mean fluctuations of density, longitudinal, and shear wave speeds. Subscripts $m$ and $c$ show the marrow and bone matrix properties, respectively. Note that this model is a weak scattering model based on small perturbations in a binary mixture in order to model the skull as a random isotropic continuum containing identical scatterers.

*2.3.3. Absorption effects.* The skull bone is a highly absorptive medium. However, in multiphase materials, such as bone, the major effect in ultrasound attenuation is due to scattering which is more than an order of amplitude bigger than absorption (Strelitzki *et al* 1998). To model the absorption effect, a power law absorption was considered here. It is well known that over diagnostic ultrasound frequencies, absorption in tissue exhibits a power law frequency dependence of the form of $\alpha = \alpha_0 f^y$, where $\alpha_0$ is the absorption coefficient, $f$ is the frequency, and $y$ is the power law exponent, typically given in range of $1 \leq y \leq 1.5$ (Treeby and Cox 2010b).

In order to incorporate the power law absorption into the wave equation, a number of approaches have been so far proposed. The most promising methods are the Chen and Holm (2003) model for lossy media and the Kramers-Kronig dispersion relation (Waters 2000, 2005). Treeby and Cox (2010b) have shown that the Chen and Holm model exhibits the desired power law absorption but is non-dispersive; thus, they have modified the absorption term of the Chen and Holm model by combining it with the Kramers-Kronig model. They applied this approach to attenuate acoustic pressure in soft tissues. This method has been used in this study for modelling the absorption effects of the skull bone. When ultrasound propagates through the skull, as an elastic medium, there will be two potentials (compressional and shear) rather one in an acoustic medium; therefore, here, each velocity potential was attenuated using the method described below. In our *k*-space method, in each time step, the potentials were attenuated according to equations (21a, b), analogous to the approach in Treeby and Cox for the acoustic *k*-space scheme,

$$\phi = \left\{1 + \frac{1}{c_p^2}\left(\tau_p k^{y-2}\frac{\partial}{\partial t} + \eta_p k^{y-1}\right)\right\}\phi \tag{20a}$$

$$\psi = \left\{1 + \frac{1}{c_s^2}\left(\tau_s k^{y-2}\frac{\partial}{\partial t} + \eta_s k^{y-1}\right)\right\}\psi \tag{20b}$$

where $\phi$ and $\psi$ represent the compressional and shear velocity potentials, and $c_p$ and $c_s$ are compressional and shear speeds of the medium, respectively. The proportionality coefficients $\tau$ and $\eta$ are given as below for each potential,

$$\tau_p = -2\alpha_{po}c_{p_\infty}^{y-1}, \tau_s = -2\alpha_{so}c_{s_\infty}^{y-1} \tag{21a}$$

$$\eta_p = 2\alpha_{po}c_{p_\infty}^{y}\tan\left(\frac{\pi y}{2}\right), \eta_s = 2\alpha_{so}c_{s_\infty}^{y}\tan\left(\frac{\pi y}{2}\right) \tag{21b}$$

The absorption coefficients, $\alpha_{po}$ and $\alpha_{so}$, are associated with compressional and shear potentials respectively, and are given by Clement *et al* (2004, 2006).

*2.4. Image reconstruction*
The image reconstruction problem in photoacoustic imaging is an acoustic inverse source problem, which is where the initial pressure distribution (i.e. the region of photoacoustic sources) is estimated from a given a set of acoustic pressure time histories recorded on a measurement surface (Treeby and Cox 2010c). For the purpose of time-reversal image reconstruction, the *k*-space forward model (which is symmetrical with respect to time) was used with zero initial conditions, but with the recorded acoustic pressure time histories used as a boundary condition at the position of the detectors. When the detectors completely enclose the initial pressure distribution, the acoustic pressure field inside the measurement surface will be re-built in time-reversed order, and the final pressure field will be the initial pressure distribution, which is the required photoacoustic image. This was implemented using the *k*-Wave toolbox facility for time reversal image reconstructions (Treeby and Cox 2010a).

**3. Results**

*3.1. Simulation layout*
To assess the photoacoustic imaging of a brain tumour, a number of simulations was set up using 3D medical images and defining a 360 degree spherical mask of ultrasound detectors. The images were obtained in the form of T1, T2 MR, Gad. MR and MRA and prepared as two segmented images, one contained only the brain tissues and the other contained both the vasculature and the tissues. They were then imported to MATLAB with the DICOM format. The images, were resized to (128×128×128) as the simulation size. A (128×128×128) space of finite element meshes with 8-noded voxel elements was generated for modelling the optical field. For the acoustic and elastic field forward solver, a (128×128×128) cubic grid domain was applied.

One hundred point detectors were randomly distributed over the head to ensure that acoustic field is spatially well sampled (see figure 5(a)). In practice, this could be realized by the provision of a helmet with embedded sensors. The algorithm for the spatial distribution of sensors was based on the Golden Section Spiral method. In the time reversal mode, these discrete data were interpolated, and accordingly, a continuous mask of data (recorded pressure) was used for the image reconstruction. The optical fibre was considered to be at position (64,25,25), in voxel coordinates. This was estimated based on the fact that in practice a possible route for accessing the cranial cavity with the optical fibre is via the nasal canal. The optical fibre was taken to be an isotropic source with a Gaussian profile. The mask of the ultrasound detectors (positioned around the head) and the position of the light source are shown in figure 5.

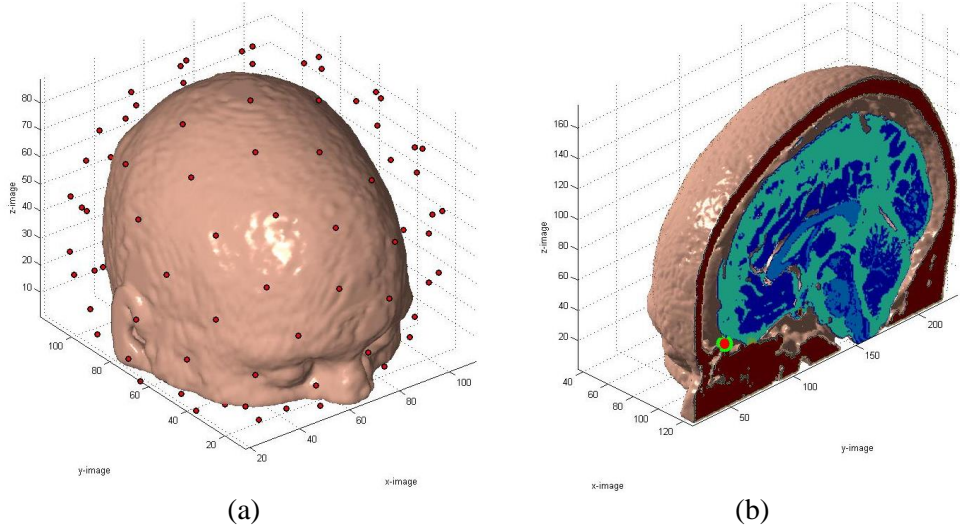

(a)          (b)

**Figure 5.** PAI simulation layout. (a) Mask of the sensors and their positions with respect to the digital phantom. (b) Position of the optical fibre with respect to the digital phantom (indicated by a red dot).

The values of the simulation parameters are presented in the appendix section. The Grüneisen parameters for brain tissues, which depicts the efficiency of tissue to convert the generated heat to acoustic pressure, are unknown. Although much smaller than that of blood cells, the optical absorption of brain tissues cannot be taken as zero, in range of the wavelengths of interest (i.e. visible light and NIR). Furthermore, since the extraction of microvasculature less than about 100 μm is not feasible when segmenting MR angiograms, each class of tissue may have a certain unresolved blood volume density after the image processing stage. Hence, the segmented vasculature may not be considered as the only photoacoustic sources.

To compensate for the former assumption and overcome the latter difficulty, the blood volume density of each class of tissue, other than the segmented vasculature, was estimated using a uniform probability distribution function for blood in each tissue type. For each class of tissue, the overall optical absorption is estimated in terms of the inherent absorption of that tissue and the blood content as,

$$\mu_{a(eff)} = \mu_{a(i)}(1 - \varphi_{bi}) + \mu_{a(b)}\varphi_{bi} \tag{22}$$

where $\mu_{a(i)}$ is the optical absorption coefficient of each class of tissue (i.e. white matter, gray matter, CSF, edema, and tumour) and $\mu_{a(b)}$ is the absorption coefficient of blood. $\varphi_{bi}$ could be considered as the volume fraction of blood for tissue $i$, and may be found in the works of Herscovitch and Raichle (1985) and van der Zee (1992).

*3.2. Effect of the skull bone on the reconstructed image*

The effect of the skull on the image was investigated by running simulations using four different models. These models are for the case of no skull, where the skull is considered as an acoustic medium, where it is an elastic homogenous medium (i.e. supporting shear wave propagation), or a highly scattering elastic medium . The scattering effect was modelled using the method described in section 2.3.2. The mask of the longitudinal velocities is shown in figure 6, where the skull is considered as purely homogeneous in 6(a) and as a highly heterogeneous medium in 6(b).

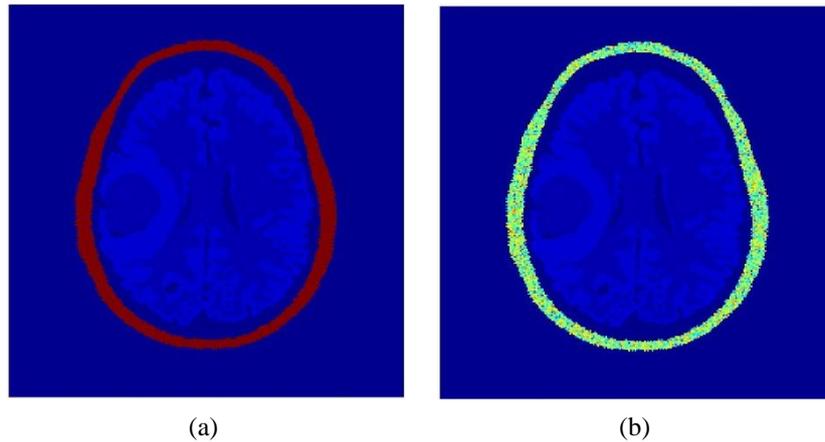

(a)          (b)

**Figure 6.** Purely elastic skull vs. scattering skull: mask of the longitudinal wave velocities. (a) Elastic skull. (b) Scattering skull.

The reconstructed images show the significance of accounting for the elastic and scattering properties of the skull. When there is no skull, the tumour is easily discernible in the images. This is also true for the case of an acoustic skull. The tumour is indistinguishable however when elastic and scattering effects of the skull are introduced.

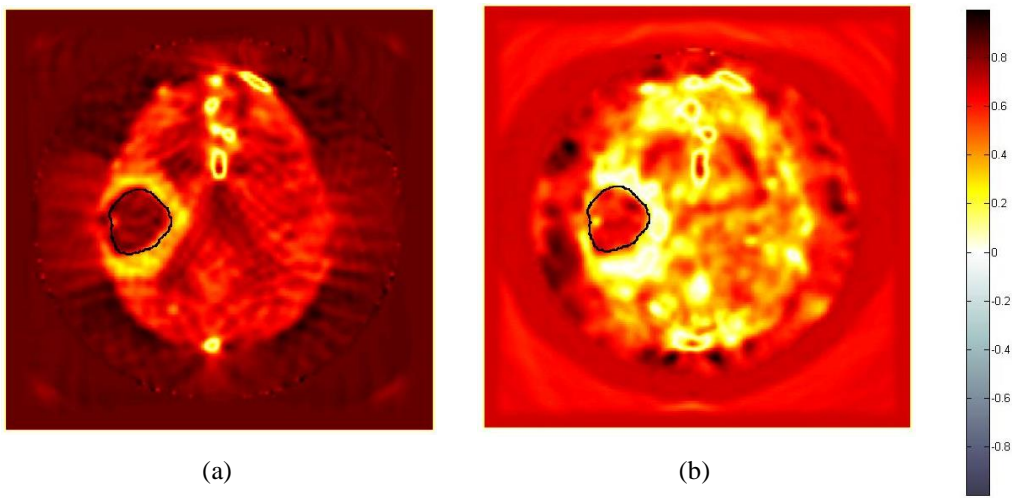

(a)          (b)

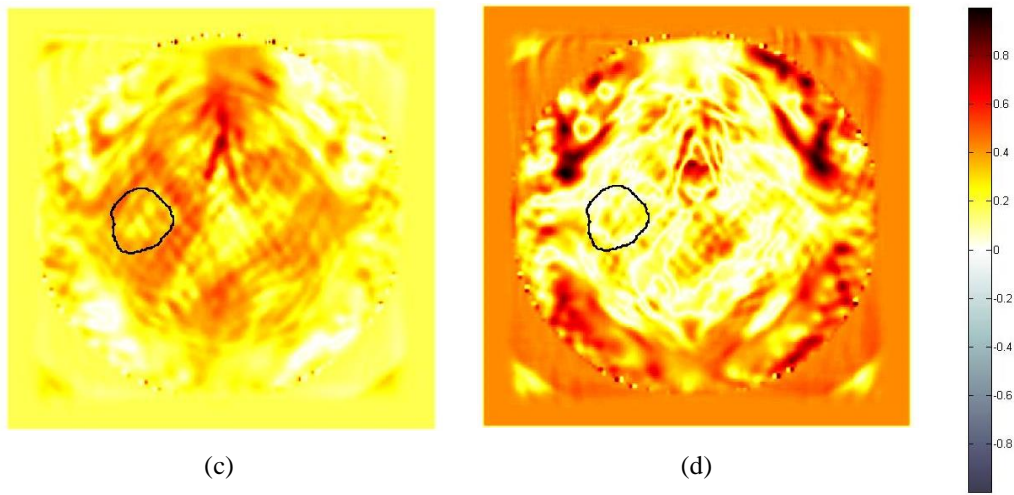

**Figure 7.** Comparison of the reconstructed images with different models of the skull (the outline of the tumour is indicated). (a) No skull. (b) Acoustic skull. (c) Elastic skull. (d) Scattering skull.

## 4. Discussion

The need for accurate modelling of the skull can be seen in the evidence presented above. To assess the quality of the reconstructed images, the border of the tumour (from the original MR scan) has been superimposed on the reconstructed images. Figure 7 shows the reconstructed images of the same simulation setup but with different skull models. Figure 7(a) is when there is no skull around the cancerous brain and as it is evident that the tumour is detectable with good contrast. Figure 7(b) illustrates the image of a brain with an acoustic skull (i.e. it supports only longitudinal waves but the skull has high contrast in speed of sound and density compared with the brain tissues). The quality of image is slightly inferior to that in 7(a). Figures 7(c) and 7(d) show the images when skull has, respectively, a purely elastic and a poroelastic nature. As it can be seen, the image quality drops dramatically in figure 7(c) compared with the previous ones and is even worse in the case of cancellous bone in 8(d). Consideration of shear wave propagation and attenuation effects of the skull leads to severe distortion of the reconstructed images. Therefore, in order to have a realistic assessment of the efficacy of photoacoustic (or ultrasound) imaging of the brain, it is paramount to use the correct model for the skull bone. Note that in the simulated images, the reconstructed pressure is based on the recorded longitudinal waves at the detectors.

Moreover, detection of the tumour highly depends on the degree of contrast of the photoacoustic properties of the surrounding tissues. Various types of brain tumours have so far been recognised. They can often be categorised in terms of the degree of abnormal growth of blood vessels throughout the tumours. They are usually hyper-vascular and some have hypo-vascular necrotic cores. A tumour should therefore produce changes in image intensity and present itself with a reasonable contrast against the surrounding healthy tissue. Here a relatively large and necrotic tumour, which has no blood content and accordingly very low optical absorption but with a vascularised surface, has been modelled. This actually has resulted in significant contrast of tumour compared with the surroundings which are vascular. This is apparent in figures 7(a) and 7(b). However, still this high contrast between blood contents has been ineffective for observing the tumour in figures 7(c) and 7(d), which is because of the significant effect of the skull on the photoacoustic field.

It needs to be emphasized that in this set of simulations both the optical and acoustic fields have been normalised. In practice however the applied optical power must not exceed a maximum permissible safe level. The safe level is characterised by MPE (maximum permissible exposure) for light fluence density, (Xu and Wang 2006).

Note that the presented results are for one particular patient only and serve to demonstrate the consistency, convergence and stability of the models used. For this case study due to the assumptions used for blood volume at the surface of the tumour (and zero at the core) the tumour cannot be clearly identified against the background.

As the next stage of this work, for a full numerical analysis of the efficacy of photoacoustic imaging for brain tumour diagnosis, the computational tool developed here can be used with a library of brain tumour MR images, containing tumours of different type, shape, size and location. It would then be possible to conduct a sensitivity analysis to help design a suitable photoacoustic imaging system for brain tumours and optimise relevant system parameters such as the position of the optical fibre source or the receiver sensor array characteristics.

## 5. Summary

A suitable numerical method was presented for the *in-silico* (simulation-based) assessment of the photoacoustic image reconstruction of the brain.

First, medical image processing (segmentation) techniques were applied to prepare a digital phantom from actual MR images. Light transport through various brain tissue types was modelled using the finite element approach. The resulting acoustic pressure was then estimated by pulsed photoacoustics considerations. The forward acoustic wave propagation (in soft tissues) was modelled by linearized coupled first order wave equations and solved by the acoustic *k*-space method.

Whereas the brain itself can be considered to be a fluid medium and does not support shear stresses, the skull bone is an elastic solid and as such is an effective shear wave propagator and strongly attenuates any ultrasound (due to both scattering and absorption). To allow the skull to support shear waves, an elastic *k*-space method was developed. To support the scattering effects of the skull, a new approach based on the theory of wave propagation in random media was applied. In addition, absorption effects were modelled using a power law. For completion of the imaging process, the acoustic pressure was reconstructed using the *k*-Space time reversal technique.

By having a complete model of the photoacoustic field, the photoacoustic tomogram of a brain tumour was generated by running a set of 3D simulations. The results depict the significance of introducing the correct skull model when assessing the photoacoustic image reconstruction. Besides, they show that the contrast of the tumour in the reconstructed images depends upon how the blood content of the tumour is defined relative to the surrounding tissues.

The results were generated for only one case in order to test the model attributes such as convergence and consistency. This now provides a suitable computation tool for further studies on photoacoustic imaging of brain tumour.

**Appendix: Table of tissue properties for the Photoacoustic Imaging (PAI) simulations**

The properties of soft tissues and the skull used for PAI simulations are given in tables (1-3).

**Tables 1.** Optical properties of soft tissues in the PAI simulations (the data are for a wavelength range of $400 - 800$nm).

| Tissue Class | Blood | Gray Matter | White Matter | CSF | Edema | Tumour |
|---|---|---|---|---|---|---|
| Scattering Coeff. (mm$^{-1}$) | 141.3 | 10.6 | 40.1 | 20 | 15 | 20 |
| Absorption Coeff. (mm$^{-1}$) | 10.2 | 0.04 | 0.08 | 0.05 | 0.1 | 0.03 |
| Refractive Index | 1.48 | 1.48 | 1.48 | 1.48 | 1.48 | 1.48 |
| Blood Volume Density (%) | - | 20 | 23 | 15 | 30 | 0 |

References: (Herscovitch and Raichle 1985, Cheong *et al* 1990, van der Zee 1992, Roggan *et al* 1999, Yaroslavsky *et al* 2002, Beard 2002).

**Tables 2.** Acoustic properties of soft tissues in the PAI simulations (the data are for a frequency range of $2 - 8$MHz).

| Tissue Class | Blood | Gray Matter | White Matter | CSF | Edema | Tumour |
|---|---|---|---|---|---|---|
| Sound Speed (m/s) | - | 1550 | 1600 | 1500 | 1580 | 1670 |
| Density (kg/m$^3$) | - | 1050 | 1030 | 1000 | 1020 | 1050 |
| Power Law Absorption Power | - | 1.1 | 1.1 | 1.5 | 1 | 1 |
| Power Law Absorption Coeff. (dB/cm) | - | 2.5 | 2.5 | 1 | 4 | 7.2 |

References: (Venrooij *et al* 1979, Bush *et al* 1993, Bamber 1997).

**Tables 3.** Properties of skull in the PAI simulations.

| Bone Matrix | | | Marrow | | | Power Law Absorption Coeff. (Np/m) | | Power Law Absorption Power | Porosity (%) |
|---|---|---|---|---|---|---|---|---|---|
| Longitudinal sound speed (m/s) | Shear sound speed (m/s) | Density (kg/m$^3$) | Longitudinal sound speed (m/s) | Shear sound speed (m/s) | Density (kg/m$^3$) | Longitudinal wave | Shear wave | | |
| 2850 | 1450 | 1900 | 2500 | 1200 | 1700 | 85 | 90 | 0.6 | 65 |

References: (Strelitzki *et al* 1998, Hayner and Hynynen, 2001, Clement *et al* 2004, 2006).